\documentstyle[epsfig,11pt]{article}
\title{\bf  Phase shift analysis in Nimtz experiments on tunneling and transmission }

\author{J.Jakiel\thanks{e-mail: jacek.jakiel@ifj.edu.pl}, W.Kantor   \\
\small {H.Niewodnicza\'{n}ski Institute of Nuclear Physics  
Radzikowskiego 152, 31-342 Cracow Poland; } }

\begin{document}

\maketitle


{\bf Abstract} For the wave representing particle traveling through any layer system we calculate appropriate 
phase shifts comparing two methods.  One bases on the standard scattering  theory and is well known another 
uses  unimodular but not unitary $M$-monodromy matrix. Both methods are not equivalent due to  different 
boundary condition - in the one barrier case there exist analytical expressions showing  difference. Authors 
generalize results to many barrier (layer) system. Instead of speaking about superluminarity we introduce into the 
quantum mechanics so called  by us "hurdling problem": can a quantum hurdler in one dimension be faster then  
a sprinter (without obstacles) at the same distance. Relations between wavefunction arguments and delay or 
advance are shown for Nimtz systems.

\section{The tunneling times definitions with reference to $S$ (scattering) and $M$ (monodromy –transfer)  
matrix theories } 
 
\setcounter{equation}{0} 
 
\subsection{Smith's method as $S$-matrix method}  
       
Before 1960 duration of a collision was a rather ill-defined concept,  depending on a more or less arbitrary choice 
of a collision distance $r$.   Such a point of view was represented by F.T.Smith (1960)[1] in his  paper "Lifetime 
Matrix in Collision Theory". In that work the author tried  to generalize delay-time $\Delta t=\hbar (\partial \eta/ 
\partial E)$  resulting from analysing the  scattering of the wave packet into a concept of the general S matrix 
theory  according to papers written by Bohm (1951)[2] and Wigner (1955)[3].  If collision time is defined  as a 
limit for $r \rightarrow \infty$, then the difference between the time: 
\\ {\bf a)} in which the interacting particle stays within distance $r$,  
\\ and the time: 
\\ {\bf b)} it would have spent there in the absence of the interaction emerges as a well-defined quantity which is 
finite if the interaction vanishes  rapidly enough at large distances $r \rightarrow \infty$ . 
 
"In quantum mechanics, using steady-state wave functions,  average time of  residence in the scattering region is 
the integrated (excess) density divided  by the total in-or out(ward) flux,  and lifetime (more precisely, time 
delay) is defined as the difference  between these residence times with and without interaction." 
 
\begin{equation} Q=\frac{\lim_{r \rightarrow \infty} \frac{1}{r} \int_{r}^{2r} 
dr'\int_{0}^{r'}(\psi^{\ast}(x) \psi (x)-\overline{\rho})dx}{f}  
= average \; integrated \; density/ flux                    
\end{equation}                    
 
where average density in the absence of the potential is: 
\begin{equation} \overline{\rho}_{\infty}=<\overline{\rho}(x)>= 
<\psi_{\infty}^{\ast}(x) \psi_{\infty}(x)>= 
\lim_{r \rightarrow \infty} \frac{1}{r} \int_{0}^{r} 
(\psi_{\infty}^{\ast}(x) \psi_{\infty} (x))dx=2AA^{\ast} 
\end{equation}                        
 
and  $j_{inw/outw}$  is the inward or outward flux as defined by Smith 
\begin{equation}  j_{inw}= AA^{\ast} \frac{\hbar k}{m} =j_{outw}= AA^{\ast} \upsilon 
\end{equation}                   
 
where, according to the scattering theory, the asymptotic (one-dimensional) form of $\psi$  at  large $x$ is: 
\begin{equation}  \psi_{\infty} (x)= A(e^{-ikx}-e^{i 2\eta}e^{ikx}) 
\end{equation}                 
and $A$ is normalization while density in the central region is: $\rho(x)=\psi^{\ast}(x) \psi (x)$ 

$\psi_{\infty} $ in case  of many channels and separable radial part of the Schr\"{o}dinger equation can be 
written as 

\begin{equation} \psi_{\infty} \rightarrow  \psi =A( \Phi^{inw}_{j} \pm \sum_{i}S_{j,i} 
\Phi^{outw}_{i}) 
\end{equation}                                  

\begin{minipage}{120mm}
\epsfig{file=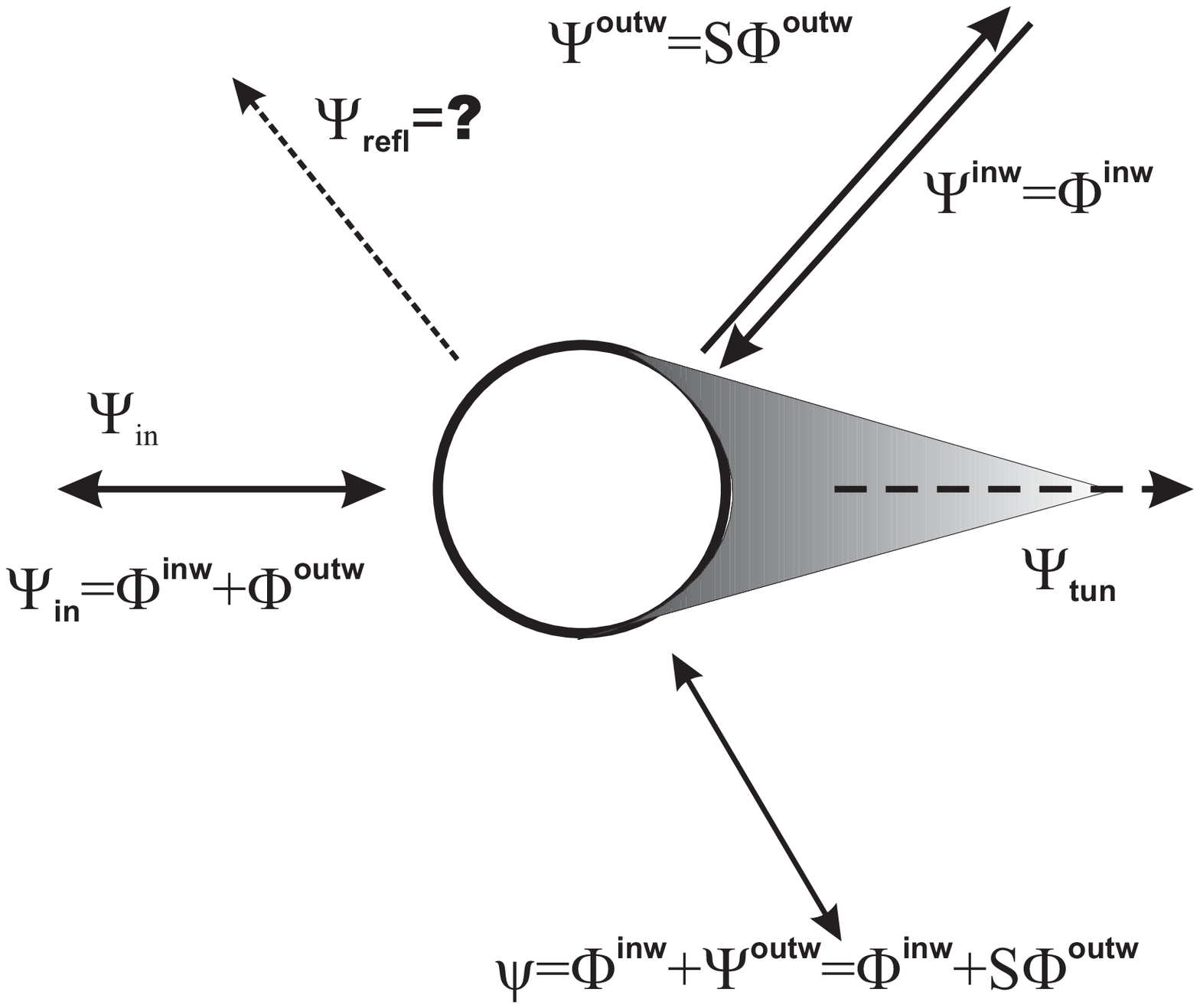,width=110mm,bbllx=50pt,bblly=240pt,bburx=530pt,bbury=650pt}
\end{minipage}

FIGURE 1. Inward,outward, tunneling etc. waves in scattering. During scattering only one wave (here) outward 
or inward is modified. \underline{ There is no cross terms between inward and outward fluxes}. $\Psi_{tun}$ is 
not incorporated in $\Psi^{outw}$ nor in $\Psi^{inw}$. It is not clear if the reflected wave $\Psi_{refl}(:|S|=1)$ 
is equal to $\Psi^{outw}$. In shadow region the complete wave function must vanish [16] ($\psi=0$), there is no 
place for $\Psi_{tun}$ in $\psi$.  

If the wave functions are normalized to inward or outward unit flux through  a sphere with  radius $r \rightarrow 
\infty$,than on the basis of  complete wave functions  (cf.fig 1)
we build  the lifetime matrix $Q$, using the time operator (there are no consistent  theory till now concerning the 
time operator )  
$t = -i\hbar \partial/ \partial E$ 
 
\begin{equation} Q=(-i\hbar \frac{\partial S}{\partial E})S^{\dag}=(tS)S^{\dag} 
\end{equation}                        
                                                                  
where $S$ is the scattering matrix. According to Smith's paper, $Q$ and $S$  contain  complementary 
information and  after diagonalization of $Q$ its eigenvalues  are the lifetimes of metastable states, while the 
corresponding eigenfunctions  are the proper functions describing these metastable states. That's why  $Q$ is  
called the lifetime matrix according to the formula derived by Smith as below: 
\begin{equation}  
 Q_{ij}=\lim_{r \rightarrow \infty} \frac{1}{r} \left[ \int_{r}^{2r} 
dr'\int_{0}^{r'}(\psi^{\ast}_i(x) \psi_j 
(x))dx-r(\frac{1}{\upsilon_{i}}\delta_{ij}+ 
\sum_{k}S_{ik}\frac{1}{\upsilon_{k}}S^{\ast}_{jk})\right]_{Av} 
\end{equation}               
where the average value is taken to eliminate oscillating terms at large r. $Q$ is introduced corollary using 
identity $Q=\hbar \partial \eta / \partial E $.

Ohmura generalized above consideration on time packets :
\begin{equation}  \psi (r,t) \rightarrow  \int  A(\omega) e^{i\alpha(\omega)}[e^{ikz}-
f(\omega)e^{i\beta(\omega)}\frac{e^{ikr}}{r}] e^{i\omega t}d\omega=\psi_{in} +\frac{1}{r}\psi_{sc}
\end{equation}  
In his method $A$,$\alpha$,$\beta$ are real functions, $\partial \alpha /\partial \omega$ gives time delay of 
incoming time packet due to reshaping before and during collision while $ \partial (\beta) /\partial \omega$ due to  
reshaping only during collision ($f^2$ is the differential cross section).  Using time dependent flux formula $j(t)$ 
averaged over time:
\begin{equation}  j= \frac{\hbar}{2im}\int_{-\infty}^{+\infty} ( \frac{\partial \psi}{ \partial r} \psi^{*}-  
\frac{\partial \psi^{*}}{\partial r} \psi)dt
\end{equation}
he got the mean time delay:
\begin{equation}  \Delta t= \frac{ \int A^2f^2(\omega)\frac{\partial (\alpha+ \beta)}{ \partial \omega }\upsilon 
d\omega}{\int A^2f^2(\omega) \upsilon d\omega}- \frac{\int A^2\frac{\partial \alpha}{ \partial \omega }\upsilon 
d\omega }{\int A^2\upsilon d \omega }
 \end{equation}

  The above idea has been applied by Olkhovsky-Racami [4] in investigations of reflection and tunneling times. 
All these methods analyze  variations of the complex wave arguments during scattering, directly (by Ohmura) 
indirectly in terms of fluxes in ref.[4]. Below we try to find analogue of phase functions $\partial \beta / \partial 
\omega$ (distributions) as function of projectile wave-number for transmission  through systems as in  Nimtz  
superluminar experiments [12].

Depending on the problem under consideration the scattered phase shifts can be defined in reference to other 
known shifts (as Coulomb phase shifts or just $ kr $ the argument of undistorted $\Phi^{inw},\Phi^{outw}$ 
waves i.e. - without interaction phase at $r$ is  given simply by $kr$ ). Now having the scatterer we replace it by 
potential (repulsive or attractive) and matching wave functions and their derivatives outside potential range (as 
solutions of corresponding wave equation with initial condition that the \underline{ wave function} is \underline 
{equal to zero} at origin as well with assumption that both fluxes inward and outward are orthogonal [1], cf. fig1) 
we calculate scattering amplitudes. The scattering device together with incoming flux is located at c.m. and 
elastic channel is one usually created by the nonresonant "reflected" wave function with the same $k$ vector. The 
scattering theory doesn't make difference between elastic reflected and transmitted waves. There is only one 
averaged elastic channel wave function. Such situation is typical for all $S$-matrix problems in area of nuclear 
reaction, the phase shifts define scattering  amplitudes and these quantities define cross sections to be considered. 
The phase shifts are not monotonic functions of energy [5] and such dependence were not investigate due to not 
unique definition of potential. There were attempts to solve the inverse scattering problem (from phase shifts to 
restore potential) but without success. 

The incoming flux when scattered by the target (barrier) is converted into the outgoing parts i.e. reflected and 
transmitted. On the projectile side in one dimension thought experiment there is reflected particle interfering with 
incident beam while on the other transmitted. But in reality it is not easy to say which particle is reflected or not. 
In the stationary theory we take into account only an averaged outgoing flux (mixture of reflected and transmitted 
particle; cf. fig1,2.). From $S$- matrix  point of  view we have in one dimension two subchannels (R,T) or as in 
case of the separable radial part of the Schr\"{o}dinger equation we must remove  l-wave degeneration. The l-
wave splits into two subfunctions corresponding to the reflected and transmitted l-wave functions. In case of the 
radial coordinate (one dimension in three dimensional space ) we are unable to define the left - right sides even 
experimentally to distinguish the reflected wave from the transmitted one (the exception is the shadow region in 
fig. 1).   This degeneracy cannot be describe with traditional $S_l$ elements.
The scattering on the set of many barriers treated as one "black box" should be described by $S$-matrix. But 
such system should be characterized by one phase shift $\eta$ or $\eta_l$ what cannot be true. We have two
functions in output each with its own phase shift. To describe such system we must introduce unimodular M-
matrix. The M-matrix conserves the mutual exclusion relation between amplitudes R and T. From the transfer 
matrix point of view each wave function should receive proper phase shift ($\varphi_R , \varphi_T$ ) after 
scattering. Then we can find $S$- matrix amplitudes if $M$- called transfer matrix is known. There is one more 
problem $S_l$ phase shifts are found from one matching while in case of transfer matrix, it is built from 
multimatching conditions as consequence of many borders between media (inside nuclear structure). The 
scattering system is no more the "black box" type. The internal structure causes multiscattering as sequence of 
subsequent reflections and refractions what requires description in terms of not unitary $M$- matrix. The 
question is if both descriptions in terms of ( $S$ if exist and $M$) are equivalent. There is no papers where $S$-
matrix R,T amplitudes could be calculated independent of $M$. ($S$ is deducted from $M$ not vice versa).

Till now the transmission (tunneling) was taken into account indirectly through reaction channels. Such treatment 
put tunneling outside the scattering theory. The consequences are :
 
-	Non unique solutions at the origin (from two solutions we take only regular one into consideration, analysis 
below shows that in tunneling case the wave function is different from zero at system origin or at least 
undefined-not used) \footnote{Jost functions start from two irregular solutions but physical meaning has their 
sum equal to zero at origin [5, ch.11, eq.(5,26,71,72)] i.e. sum is regular .}. 

-	In nuclear physics there are problems with hard or soft core potentials which were not tested or compared 
with tunneling.

-	Validity of time reversal invariance or detailed balance theorem which says that the time reversed  incoming 
state (under the operator $\hat{K}$) is equal to an outgoing state with the same energy. The reversed in-state 
goes into  the asymptotic free time reversed state $\Phi_{i'}$ when $t \rightarrow +\infty$ i.e. 
$\hat{K}\Psi^{(+)}_{i}=\Psi^{(-)}_{i'}$ and $\hat{K}\Psi^{(-)}_{f}=\Psi^{(+)}_{f'}$. These relations 
induce $S_{i'f'}=S_{fi}$ called detailed balancing or microreversibility. In other words the transition 
probability for the inverse process with time-reversed parameters is the same as that of the direct process. 

But tunneling is irreversible process and –we suspect- cannot be described by function regular at origin.
 
In general case of the  reaction $a+A \rightarrow b+B$ (in the subbarrier collision) tunneling in out-state  
($b+B$) is different from that in in-state ($a+A$) and tunneling disturbs scattering states.  Clearly nonunitary 
condition breaking microreversibility relation , tunneling however can be introduced as additional 
indeterminance in scattering  theory.  

\subsection{ Weak Wigner causality and Wigner time }

The Wigner time is the simplest one. According to [3] and formulated there the principle of causality, the 
scattered wave cannot leave 
the scatterer (of diameter r) before the incident wave has reach it  i.e. $ \partial \eta / \partial k > -r $. This 
expression in case of positive derivatives gives retardation while negative values – advanced in time solution, for 
the outgoing wave as defined in [3] we can write
$ t_{out} = \frac{r}{\upsilon}+\frac{2\partial \eta}{\upsilon \partial k}$. Experimentally it is not easy to find 
from the excitation functions (cross sections) $\eta (k)$.(cf.eq. 1.10). In reality in macro world the scatterer 
(Coulomb or gravity field) has infinite radius what forces $ t_{out} \rightarrow \infty$. 
Let  $d=2r$ and  $t_{in}=- \frac{r}{\upsilon}$ then 
\begin{equation}  t_{Wigner}= t_{out}-t_{in}= \frac{d}{\upsilon}+\frac{2\partial \eta}{\upsilon \partial k}
\end{equation}

 If we know $\eta (k)$ the Wigner time (the group one) can be easy derived for finite systems.

\subsection{ Monodromy}

To introduce $M$ matrix we need two ingoing $in(\pm )$ and two outgoing $out(\pm )$ particle wave functions, 
(cf.Fig2).

FIGURE 2. {The monodromy problem as defined in [6,10] for one barrier. Mutual relations between ingoing and 
outgoing  (from left or right side) particle wave functions are displayed. (In the picture bars mean complex 
conjugations.) The transition from initial state  to final one $\textstyle{ \left( {\begin{array}{*{20}c}
   {\Phi _{( + )} }  \\
   {\Phi _{( - )} }  \\
\end{array}} \right)_{in}  \to \left( {\begin{array}{*{20}c}
   {\Phi _{( + )} }  \\
   {\Phi _{( - )} }  \\
\end{array}} \right)_{out}} $ is given by the unimodular M matrix  $\textstyle{ \left( {\begin{array}{*{20}c}
  {1/\bar T} &    {\bar R/\bar T} \\
{R/T}  & {1/T}   \\
\end{array}} \right)} $ easy deducted from transitions as drawn in picture. }

\begin{minipage}{160mm}
\epsfig{file=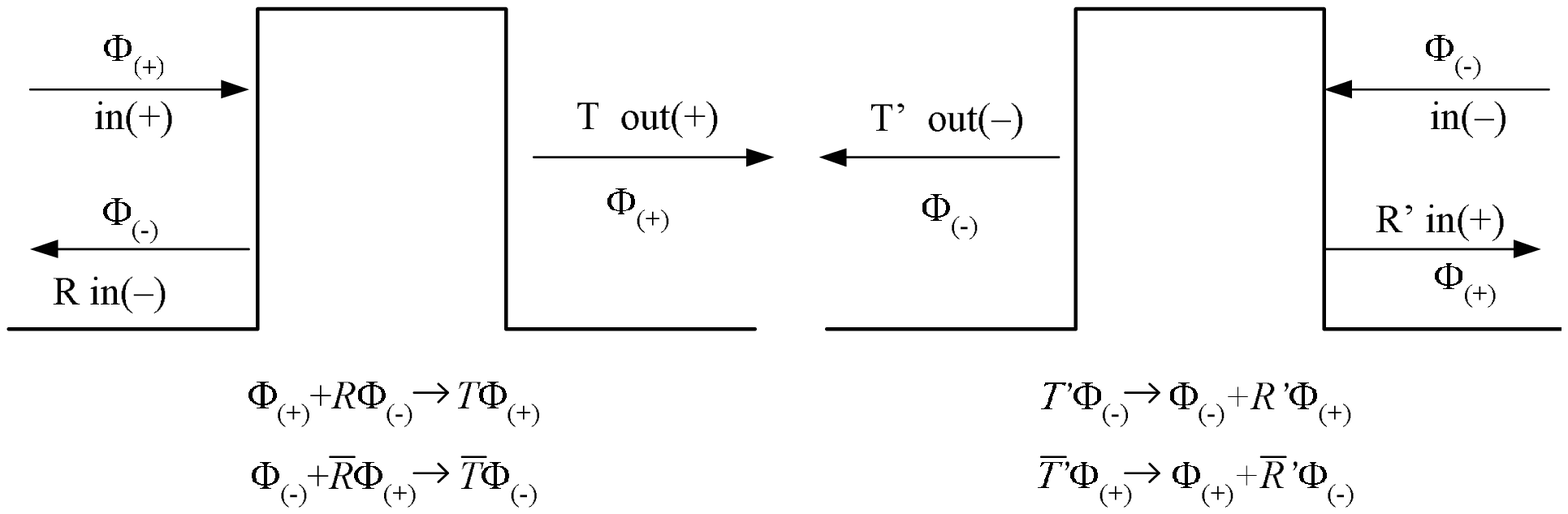,width=150mm,bbllx=0pt,bblly=10pt,bburx=590pt,bbury=280pt}
\end{minipage}

The transmission or reflection  through any periodic or aperiodical set of square barriers rewritten as the 
transformation from $(\Phi_{in} , \Phi_R)$ to $(\Phi_T)$ (undergoing the monodromy matrix), can be described 
in terms of  the $[2 \times 2]$transfer ("monodromy") cells given by superposition of  $[O^{\pm}_i]$ matrices  
with $[H_i]$ . $[O^{\pm}_i]$ represent free wave propagation between barriers which can be interpreted as 
phase translation to given position (the middle and/or edge of barriers) or phase translation about relative 
distance if  $[O^{\pm}_{\Delta (i)}]$ are used . In ref. [7] authors use simply the name translation operator. 
$[H_i] $ describe particle motion under any barrier and are responsible for interactive wave propagation. 
Superposition of both matrices $[H_i] [O_i]$   represents propagation in two opposite directions inside or 
outside media. In case of the square barriers unimodular $[H_i] $ matrices have form 
$ \left[ {\begin{array}{cc}   a  & b  \\  c & a   \end{array} } \right] $ where $a,b,c$ are real.

Monodromy (or the transfer matrix or translation operator unimodular not unitary (as well not equivalent to 
unitary) transforms the initial wave function amplitudes $A_0,B_0$ or  $(1,R)$ onto outgoing one $ A_n,B_n$ 
or $ (T,0)$. 

\[
\left[ {\begin{array}{*{20}c}
   A_n  \\
   B_n  \\
\end{array}} \right] =\mathcal{ M} \left[ {\begin{array}{*{20}c}
   A_0  \\
   B_0  \\
\end{array}} \right]  
\]

The monodromy form of $\mathcal{M}$ depends on the basic wave functions to be chosen. $M$ shifts the 
solution of  the Schr\"{o}dinger equation from $x$ to $x+d$ i.e. from beginning of the barrier system to its end. 
In the time depended approach the wave function underlies unitary evolution :
$\Psi(t=+\infty)=U(+\infty,-\infty)\Phi_{in}(-\infty)$. The initial and final wave functions are separated in time in 
$S$-matrix treatment while in $M$ matrix approach as well in space : far left, far right.

For inward, outward [(complex exponential functions, Hankel functions etc) or real basic solution like ($\cos , 
\sin $, regular and irregular Coulomb or Bessel functions etc.) , here both representation $M$ and $M'$ are 
connected by the unitary transformation] we get: 

\begin{equation}
\begin{array}{lr}
 \left[ {\begin{array}{*{20}c}
   T  \\
   0  \\
\end{array}} \right] = M\left[ {\begin{array}{*{20}c}
   1  \\
   {R}  \\
\end{array}} \right]  ;   & 
\left[ {\begin{array}{*{20}c}
   T  \\
   {iT}  \\
\end{array}} \right] = M' \left[ {\begin{array}{*{20}c}
   {1 - R}  \\
   {i(1 + R)}  \\
\end{array}} \right] \\ 
 \end{array}
\end{equation}

Monodromy $M$ represents propagation of the wave functions through system of multiple cells as sequence of 
reflections and transmission  (at each cell edge the wave is splitted into reflected and refracted (tunneled)) or 
$M$ can be interpreted as superposition of  cells characterized by two waves inward and outward. The four $M$-
matrix elements can be expressed as function of complex variables $T$ and $R$, above relations define only 
$M_{21}= R/T$ , $M_{22}=1/T$ elements. The remaining $M_{12}$ , $M_{11}$ elements, connected by 
$det[M]=1$ relation, we deduce from matching conditions. The monodromy is unimodular not unitary. $M^{-
1}\ne M^{\dag}$ and hermitian conjugation does not describe inverse motion.  Multiple reflections and 
transmissions are strictly correlated with multiple matching. If system is asymmetric (there exist at least one left 
and right matching which do not coincide), equivalent $S$-matrix can exist if we introduce additional phase shift 
$\Delta \varphi$ between $R$ and $T$ waves. Monodromy can be periodic but not necessary. Without 
dissipation (energy loss) system consist of multiple superposition of unimodular matrices.

Now we consider transmission through certain device created by superposition of  many barriers. Such systems 
can be equivalent any arbitrary shape potential $U (x)$ defined on the intervals $a_i - \epsilon_i \le x \le a_i + 
\epsilon_i $ with help of square barriers (e.g. barriers on the Cantor set etc.).
There are barriers as in the fig.3:

\begin{minipage}{100mm}
\epsfig{file=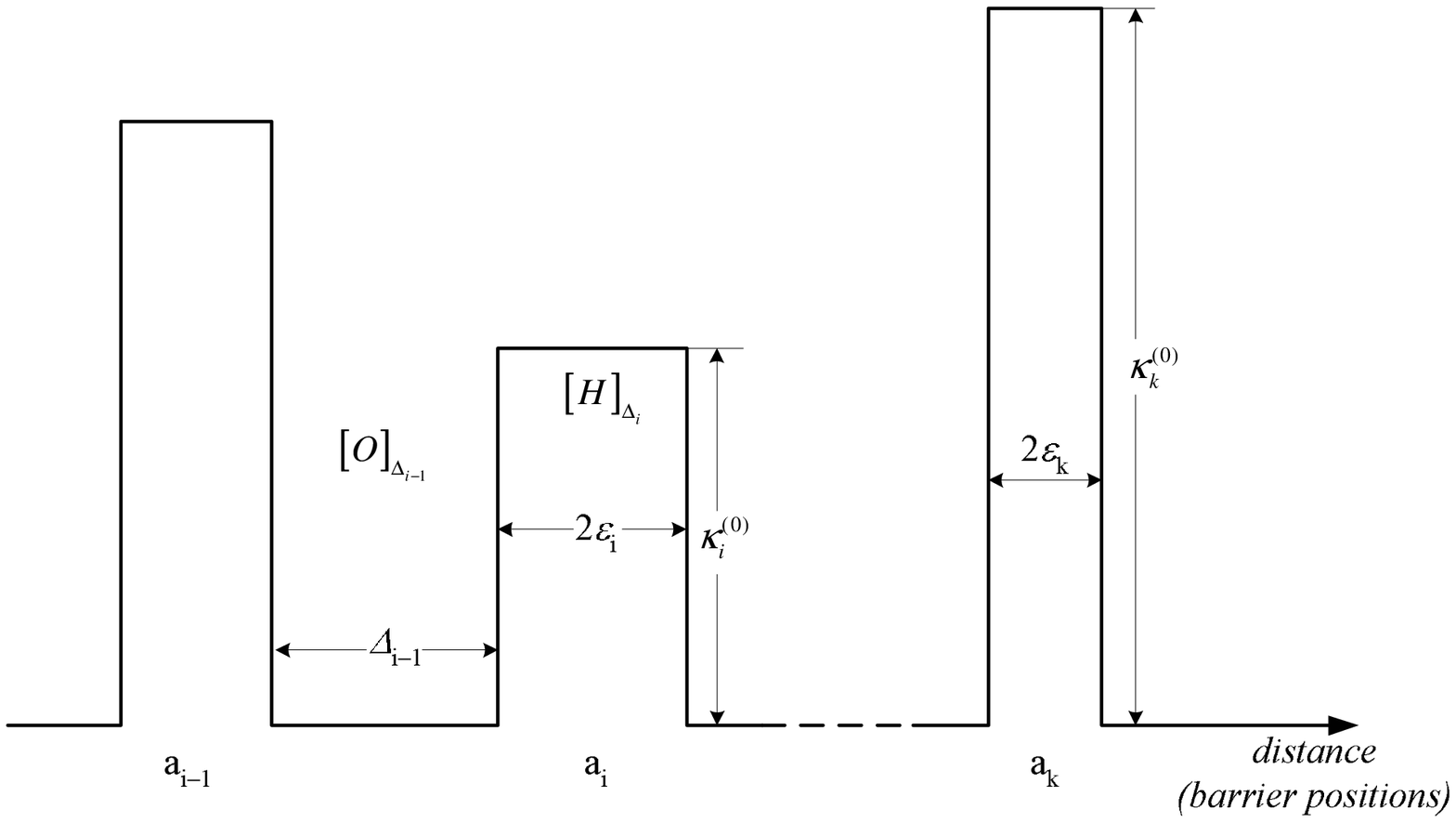,width=90mm,bbllx=0pt,bblly=0pt,bburx=530pt,bbury=350pt}
\end{minipage}                                             

FIGURE 3.     Set of barriers as used in tunneling and transmission or reflection.

The superposition of all  "phase translations" describe the full transfer operator $M$ as transformation from the 
initial (spinor) amplitude state to final one by means of matching conditions. By appropriate unitary 
transformation we can choose convenient amplitude representation. The choice depends on physics to be 
considered.

The n-th barriers system can be described by barrier center coordinates ($a_j$) or interbarrier distances 
($\Delta_j$) and  barrier widths and heights  ($2\epsilon_j , \kappa_j^{(0)}$). Then  the system width is given as

\begin{equation}
\begin{array}{lcr}
d = \varepsilon _n  + a_n  - a_1  + \varepsilon _1   &  or  & d = \sum\nolimits_j {(2\varepsilon _j  + \Delta _j )} \\   
a_{k+1} = \Delta a_k +a_k   &   & = \sum\nolimits_j {d_j }  \\ 
\end{array}  
\end{equation} 

where $a_i$ is the $i$-barrier center position and $2\epsilon_i$ width of the $i$-th barrier, $ \Delta a_k$ - 
interbarrier distance (between barrier centers) and  $\Delta _j  $ free cell width (distance between neighbour 
barrier edges.

The $M$ transfer matrix can be expanded as multiplication of  $\left[ {H_j } \right]  $and $\left[ {O_j^{( \pm )} 
} \right] $ matrices. The first discribe particle motion under the barrier (or in media) while second free motion 
between barriers. 
The matrices ${ [O]_j }$ can be written as the function of $\Delta_j$ i.e. distances between 
adjacent barriers given by the difference of their edge positions:

\begin{equation}
\Delta _j  = (a_{j+1}  - \varepsilon _{j+1} ) - (a_{j }  + \varepsilon _{j} )
\end{equation} 

The transfer matrix can be expressed in terms of the barriers edge i.e. $a_i \pm \epsilon_i$ or distances between 
adjacent barriers eq.(1.13)  then the transfer operator ${M'}$ is

\begin{equation}
[M'] = \left[ {H_n^{} } \right] \prod\limits_{i = 1}^{n - 1} {\left[ O_{\Delta _i }\right]  \left[ H_i \right] } 
\end{equation}

In representation such that $ \left[ O_{\Delta _i }\right]  $ is diagonal,  we have the final form of the position 
independent complex transfer operator 
 
\[
U_M^{} \left[ {M'} \right]U_M^ {\dag}     = \left[ M \right]
\]
in case of the  $ \cos , \sin $ base  $U$ is the unitary matrix which make diagonal   $ \left[ O_{\Delta_ i } \right] $ 
i.e 
   

\[
\left[ O^{(U)}_{\Delta_j} \right] =  
U_M \left[ O_{\Delta_j} \right]  U_M^ {\dag }    = \left[ {\begin{array}{*{20}c}
   {e^{ik\Delta _j } } & 0  \\
   0 & {e^{ - ik\Delta _j } }  \\
\end{array}} \right]
\]
$\left[ O^{(U)}_{\Delta_j} \right] $ is interpreted as stream of two waves propagating in opposite directions.

All that matrices, denoted by $ \left[ {\mathcal M^{(U)}} \right] $, like $\left[ H_n^{(U)} \right] $, $\left[ M 
\right]$, and  $\left[ O \right]$ after diagonalization, belong to monodromy type [6] ,(cf fig2.) i.e.

\begin{equation}
 \left[ {\mathcal M^{(U)}} \right] = \left[ {\begin{array}{*{20}c}
   {X + iY} & { - V - iW}  \\
   { - V + iW} & {X - iY}  \\
\end{array}} \right]
\end{equation}

 If $\left[ {M'} \right]$ is real and $ \left[ {M} \right] = U_M^{} \left[ {M'} \right]U_M^ {\dag} $
then  $ X=(M'_{11}+ M'_{22})/2 $; $W=(M'_{11}- M'_{22})/2 $; $ V=(M'_{12}+ M'_{21})/2 $; 
$Y=(M'_{12}- M'_{21})/2 $; 

V and W gives information about asymmetry in tunneling (breakdown of balance between two waves traveling 
inside the "black box" in two opposite directions).

The product of $\left[ O^{(U)}_{\Delta _i} \right] \left[ H ^{(U)}_{i}\right] =M_i $ is then the element of the 
barrier structure named the single cell transfer operator. It can be written as:

\begin{equation} 
\begin{array}{l}
 M_i  = \left[ O_{\Delta _i }^{(U)} \right]  H_i^{(U)}  = 
 \left[ {\begin{array}{*{20}c}
   {e^{ik\Delta _i } } & 0  \\
   0 & {e^{ - ik\Delta _i } }  \\
\end{array}} \right]  \\ 
\left[ {\begin{array}{*{20}c}
   {\cosh (2\kappa _i \varepsilon _i ) + {\textstyle{i \over 2}}({\textstyle{1 \over {\sigma _i }}} - \sigma _i )\sinh 
(2\kappa _i \varepsilon _i )} & { - {\textstyle{1 \over 2}}({\textstyle{1 \over {\sigma _i }}} + \sigma _i )\sinh 
(2\kappa _i \varepsilon _i )}  \\
   { - {\textstyle{1 \over 2}}({\textstyle{1 \over {\sigma _i }}} + \sigma _i )\sinh (2\kappa _i \varepsilon _i )} & 
{\cosh (2\kappa _i \varepsilon _i ) - {\textstyle{i \over 2}}({\textstyle{1 \over {\sigma _i }}} - \sigma _i )\sinh 
(2\kappa _i \varepsilon _i )}  \\
\end{array}} \right] \\ 
 \end{array}
\end{equation}

where $ \sigma_i = \textstyle{\kappa_i \over k } $ and $ \kappa_i^2 =(\kappa_j^{(0)})^2 - k^2 $, k is the 
projectile momentum. 
Transition to $\delta$ barrier set is done using eq.(1.17) when in ${ M_i}$ we put $ \lambda =2 \kappa^2_i 
\epsilon $ then we get:
 
\[
M_j^{(\delta )}  = \left[ {\begin{array}{*{20}c}
   {e^{ik\Delta _j } } & 0  \\
   0 & {e^{ - ik\Delta _j } }  \\
\end{array}} \right]\left[ {\begin{array}{*{20}c}
   {1 + i{\textstyle{{\lambda _j } \over {2k}}}} & { - {\textstyle{{\lambda _j } \over {2k}}}}  \\
   { - {\textstyle{{\lambda _j } \over {2k}}}} & {1 - i{\textstyle{{\lambda _j } \over {2k}}}}  \\
\end{array}} \right]
\]

using two phase representation (of reflection and tunneling )[8] we can rewrite the ${ [M_i] }$ matrix for 
symmetric structure in more compact form:

\begin{equation}
\left[ {M_i } \right] = \left[ {\begin{array}{*{20}c}
   {e^{ik\Delta _i } } & 0  \\
   0 & {e^{ - ik\Delta _i } }  \\
\end{array}} \right]\left[ {\begin{array}{*{20}c}
{\frac{{e^{i(\varphi _{1,i}+\varphi _{2,i}) } }}{{\sin (\varphi _{1,i})}}} & { - \cot(\varphi _{1,i}) e^{ i\Delta 
\varphi_{i} }}  \\
   { - \cot(\varphi _{1,i}) e^{- i\Delta \varphi_{i} }} & {\frac{{e^{ - i(\varphi _{1,i}+\varphi _{2,i}) } }}{{\sin 
(\varphi _{1,i})}}}  \\
\end{array}} \right]
\end{equation}  
where we put on the base of the single barrier transmission formula :

\begin{equation}
\varphi_{Monodromy,i}=\varphi _{1,i}+ \varphi _{2,i}= \tan^{ - 1} ({\textstyle{1 \over 2}}({\textstyle{1 \over 
{\sigma _i }}}-\sigma _i)\tanh (2\kappa _i \varepsilon _i ))
\end{equation} 
 reseparation of both phases in general case (many barrier system) is not easy.

This single cell operator can also be written with help of amplitudes ($R_{1,i},T_{1,i}$) and phases 
($\varphi_{2,i}, \varphi _{1,i} $) as:

\begin{equation} 
\addtocounter{equation}{-2}
M_i  = \left[ {O_{\Delta _i }^{(U)} } \right]H_i^{(U)}  = \left[ {\begin{array}{*{20}c}
   {e^{ik\Delta _i } } & 0  \\
   0 & {e^{ - ik\Delta _i } }  \\
\end{array}} \right]\left[ {\begin{array}{*{20}c}
   { \frac{1}{ {T_{1,i}^{\dag} }}e^{ ( i\varphi _{2,i} + \varphi _{1,i})} } & { - \frac {R_{1,i}^{ \dag} }
{ T_{1,i}^{\dag}} e^{ i\Delta \varphi_{i} } } \\
   { - \frac{{ R_{1,i} }}{{ T_{1,i}}}e^{-i\Delta \varphi _{i} } } & {\frac{1}{{ T_{1,i}}}e^{-i(\varphi _{2,i}   + 
\varphi _{1,i})} }  \\
\end{array}} \right]
\end{equation}

We assume $ T_{i}= T_{1,i}e^{ i(\varphi _{2,i}+ \varphi _{1,i}) }$ and $R_{i}=R_{1,i}e^{i(\varphi _{2,i}+ 
\varphi _{1,i}+\Delta \varphi_{i})}$
\footnote{Due to relation $|R|^2+|T|^2=1$ all cyclic function can be calculated modulo $\pi$ i.e $T_{1,i}=| 
T_{1,i}|e^{\pm i\pi}$ (real (cos)sinusoidal amplitude$=\pm$ modulus) while the phase between R and T 
amplitudes due to  imaginary factor $i$ modulo $\pi/2$ (equivalence of $\tan(nx)$ and $\cot(nx)$ sets, see [2] 
ch.11, Bohm's $\varphi_i$ are different from ours, his $\varphi_{refl}=\varphi_{trns}\pm \pi/2$)  ; we require 
only smooth behavior of phase function and its derivative.}  
So the wave phase in the reflection channel differs by $ \Delta \varphi_{i}$ from transmitted one.
$\Delta \varphi_{i}$ can be computed from the expression 
 
\begin{equation}
\addtocounter{equation}{1} 
\tan (\Delta \varphi ) = \frac{W}{V}
\end{equation}
$\Delta \varphi \neq 0$ for asymmetric systems, for symmetric one we can put $\Delta \varphi = 0$  

\subsection{ Phase properties of $M$ matrix-one barrier case or the barrier set equivalent to one }

Let in the equation (1.17,18) $\Delta = const$,$\epsilon_i = \epsilon$ and $ H$ is the same for all $i$ $H_i 
=H_{\epsilon}$. Then we call $M_i = M_{\epsilon, \Delta} $ single cell "power"(or periodic) monodromy 
operator.  Internal structure of $[M]$ representing certain device causing reflection and refraction defines 
transmission or tunnelling through the barriers as well as general (aperiodic) monodromy. We maintain that 
monodromy as applied to two channel elastic scattering is group property .

\qquad Let we make one more comment: periodical structure emerge with multiple application of  $ \left[ 
O_{\Delta}  \right] H_{\epsilon} $ but final boundary condition can change periodicity.

We can solve eq.(1.12) with $M$ as in (1.16,17) to find amplitudes $R$ and $T$

\[  iT e_{n}^{ ik(a_{n} + \epsilon_{n})  }   = M^{}_{11} ie_0^ { ik(a_{1}-\epsilon_{1})}  + M^{}_{12} e_0^ 
{- ik(a_{1}-\epsilon_{1})}  R   \]
\[ 0 = M^{}_{21} ie_0^{ ik(a_{1}-\epsilon_{1})}   + M^{}_{22} e_0^{ -ik(a_{1}-\epsilon_{1})}  R  \] 


The general solution of  that equation  is:

\begin{equation}
T = \frac{{M_{11}^{}  M_{22}^{}  - M_{12}^{}  M_{21}^{} }}{{M_{22}^{}}}e^{ - ikd}  = \frac{1}{{X - 
iY}}e^{ - ik(a_n  + \varepsilon _n  - a_1  + \varepsilon _1 )}  
\end{equation} 

\begin{equation}
R = \frac{{-W + iV}}{{X - iY}}e^{  2ik(a_1  - \varepsilon _1 )}  
\end{equation}
 
We can calculate also the amplitude ratio 
  
\[
\frac{{R^{} }}{{T^{} }}  = {\rm }  \frac{{R_1^{} }}{{T_1^{}}}\exp (i(\Delta \varphi  + k(a_n  + \varepsilon 
_n  + a_1  - \varepsilon _1 ))){\rm  } 
\]

In above formulas we put

\[ 
 M^{}_{22}= X-iY= \frac {\exp(- i(\varphi _1 +\varphi_2))}{T_1}= \frac {\exp (- i \varphi_{Monodromy}) 
}{\sin (\varphi _1 )}
\]  
and system total width $d= a_n  + \varepsilon _n  - a_1  + \varepsilon _1$ can be expressed by interbarrier 
spacing $\Delta_i$ i.e.: $d= \sum_{i} (2\epsilon_i +\Delta_i)= \sum_{i} d_i $. In that way $d_i$ defines the 
single cell width (one barrier plus one interbarrier well). 
As we have seen in explicit formulas for $M^{}_{11} $ and $M^{}_{22}$ the argument of diagonal matrix 
elements is $ \arg(M^{}_{11}) = \varphi _1 +\varphi_2  =\varphi_{Monodromy} $ and it depends through 
$\varphi _i $ on internal system structure.

Usually for symmetric systems we have  $W=0$ then $V $ is real and both amplitudes have the same phase.

The expression (1.21,22) for amplitudes $R$,$T$ depend on the wave function value  at the "black box" edges.  
In case of symmetrical aperiodical systems  we put $a_n +\epsilon_n =-a_1 +\epsilon_1 $. 

For one barrier using unimodularity, in case of $R$ and $T$ we get:

\begin{equation} 
\addtocounter{equation}{-2}
iT = \frac{i}{{M^{}_{22} }}e^{ - ikd}  = i\sin (\varphi _1 )\exp \left[ {i ( \tan^{ - 1}{\textstyle{1 \over  
2}}({\textstyle{1 \over {\sigma _1 }}}-\sigma _1)\tanh (2\kappa _1 \varepsilon _1 )-kd) } \right]
=i\sin (\varphi _1 )\exp (i\bar{\varphi}_2) 
\end{equation} 
 
\begin{equation} 
R = \frac{{ - M_{21}^{}  ie^{2ik(a_1  - \varepsilon _1 )} }}{{M_{22}^{} }} = \cos (\varphi _1 )\exp \left[{i ( 
\tan^{ - 1}{\textstyle{1 \over  2}}({\textstyle{1 \over {\sigma _1 }}}-\sigma _1)\tanh (2\kappa _1 \varepsilon _1 
)-kd) } \right]= \cos (\varphi _1 )\exp (i\bar{\varphi}_2)
\end{equation} 
We can rewrite  $\bar{\varphi}_2$ as follows 
\begin{equation}
\bar{\varphi}_2= -kd+2\tan^{-1}{(\frac{1}{\sigma_1}\tanh (\kappa_1 \epsilon_1))} +\tan^{-
1}\frac{2\sigma_1}{(1+\sigma^2_1)\sinh (2\kappa_1 \epsilon_1) }=2\eta+\varphi_1
\end{equation}

Here $\eta$ is as in [9] (Aufgabe 57) i.e. $\eta=-kd/2+\tan^{-1}{(\frac{1}{\sigma_1}\tanh (\kappa_1 
\epsilon_1))} $ , Fl\"{u}gge to solve problem put $\psi (0)=0$, we do not need that condition and "translational" 
boundary conditions at $r=r_{0}\pm d/2$ (assuming $r_{0}=0$) result in additional phase $\varphi_1$. Then  in 
our methods cross section is proportional to $\sin^2(\bar{\varphi}_2/2)$ not $\sin^2(\eta)$ . 

In  fig. 4,5 we have shown transmission , $\delta\arg{T}=\bar{\varphi}_2(k)$ and $\varphi_2(k) =2\eta+kd$ in 
single barrier case. The  
$\bar{\varphi}_2(k)$ in allowed k-band is increasing function of k and the quantum hurdler is slower then 
particle without obstacle; the $\eta(k)$ function has not such properties.

Using $\varphi_{Monodromy} - Et=0$, from position of the packet center we find the transmission time in the 
Nimtz experiments. 
 
The heuristic time calculation fulfills typical limits :

\[ 
\frac{{\hbar \partial (\varphi _{1,i}+ \varphi _{2,i})}}{{i\partial E}}\mathop  \Rightarrow \limits_{\kappa 
\varepsilon  \to \infty }- \frac{2}{{\upsilon \kappa_i }}
\]
For typical $\varphi_2$ expressions we get the same limit. That expressions are not here important. We presume    
wide systems are composed from thin elementary segments.

FIGURE 4. Transmission through one barrier three units [mm] wide. $\kappa^{(0)}$ is barrier height in $k$ 
units [1/mm]. $-kd$ represents maximal negative phase slope according to weak Wigner causality. Slope of  
$\bar{\varphi}_2=\varphi_2 + \varphi_1-kd$ is connected with the group velocity in transmission through the 
barrier. $(\partial/\upsilon\partial k) (\bar{\varphi}_2)=\delta \tau$ is time "delay". The wave length $\lambda$ at 
$\kappa^{(0)}$  is smaller than the barrier width.

\begin{minipage}{100mm}
\epsfig{file=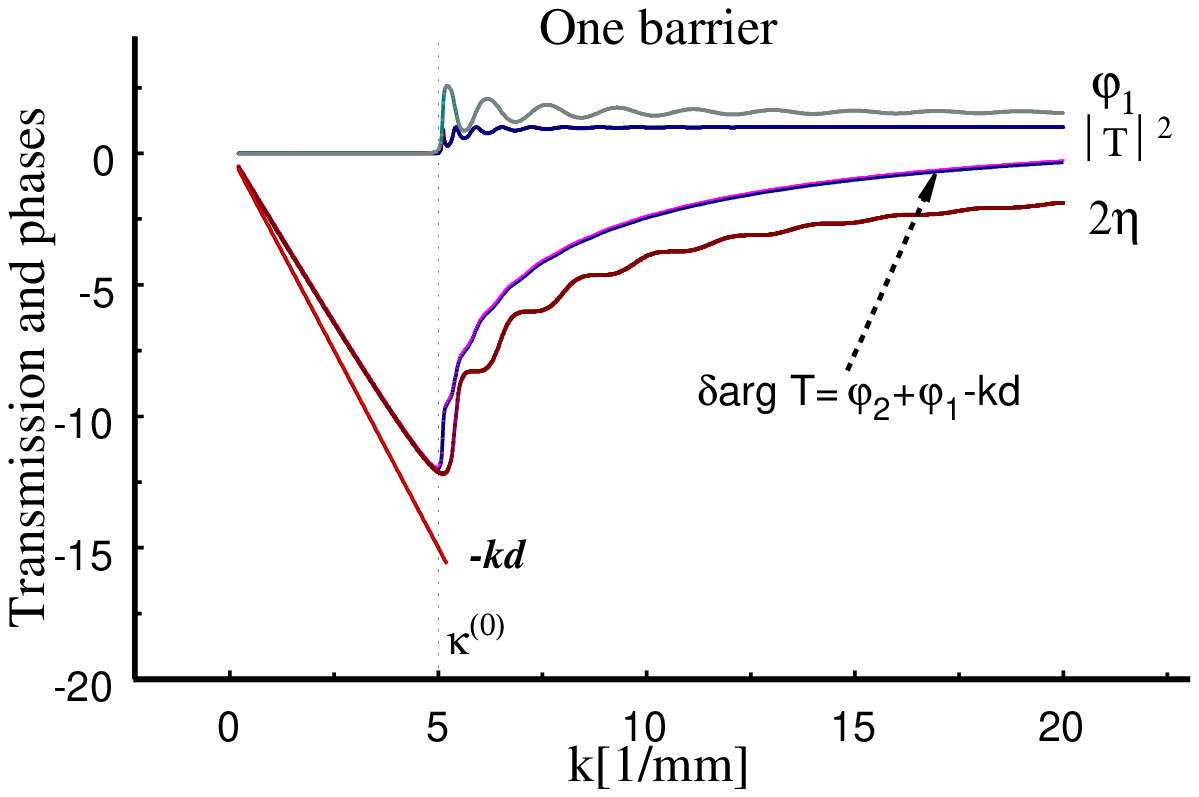,width=90mm,bbllx=110pt,bblly=80pt,bburx=470pt,bbury=340pt}
\end{minipage}                                             

FIGURE 5. Transmission through one Nimtz barrier 6mm wide. $\kappa^{(0)}$ is barrier height in $k$ units 
(here $mm^{-1}$). As in fig.4 ($-kd$) represents maximal negative phase slope according to weak Wigner 
causality. The wave length $\lambda$ at $\kappa^{(0)}$  is bigger than the barrier width so phase characteristic 
is dominated by $\varphi_1$. The slopes of $\varphi_1(k)$ and $\delta \arg{T}$  phase curves  are positive and 
give retardation. However $\varphi_2-kd=2\eta$ suggests small speed advance.

\begin{minipage}{100mm}
\epsfig{file=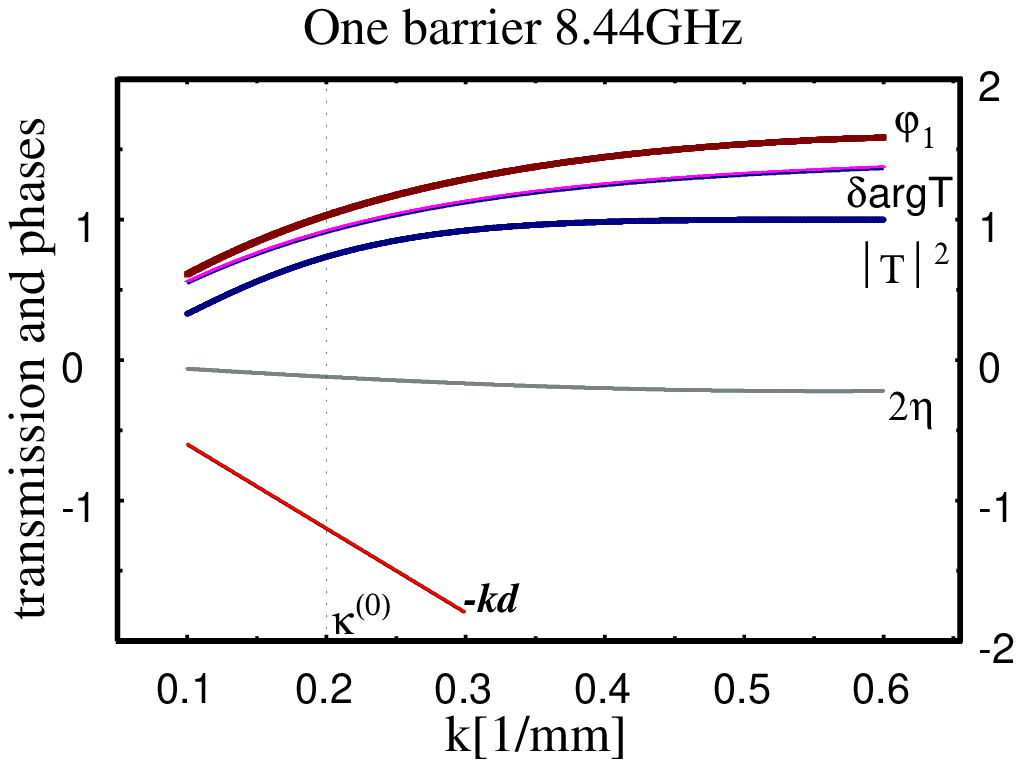,width=90mm,bbllx=110pt,bblly=110pt,bburx=420pt,bbury=350pt}
\end{minipage}

In Fig.4 we have shown the phase characteristics in tunneling through one (or two barriers see fig.6). Due to 
weak Wigner causality applied to the sum of both phases  $(\partial/\partial k) (\bar{\varphi_2}) > -d $. From 
monodromy $\delta \arg{T} =\delta \arg{R} =\varphi_2 + \varphi_1-kd=\bar{\varphi}_2 $. The phase $\varphi_2 
-kd$ alone is typical $2\eta_0$ as in Q.M.-textbooks (see [5,9]). Sometimes for -one or  few barriers  - when 
$\lambda$ is comparable with the barrier width, the height of the barriers can be easy deduced from the phase 
characteristics. It is not a rule cf. fig 5, 7 and others.

\begin{minipage}{100mm}
\epsfig{file=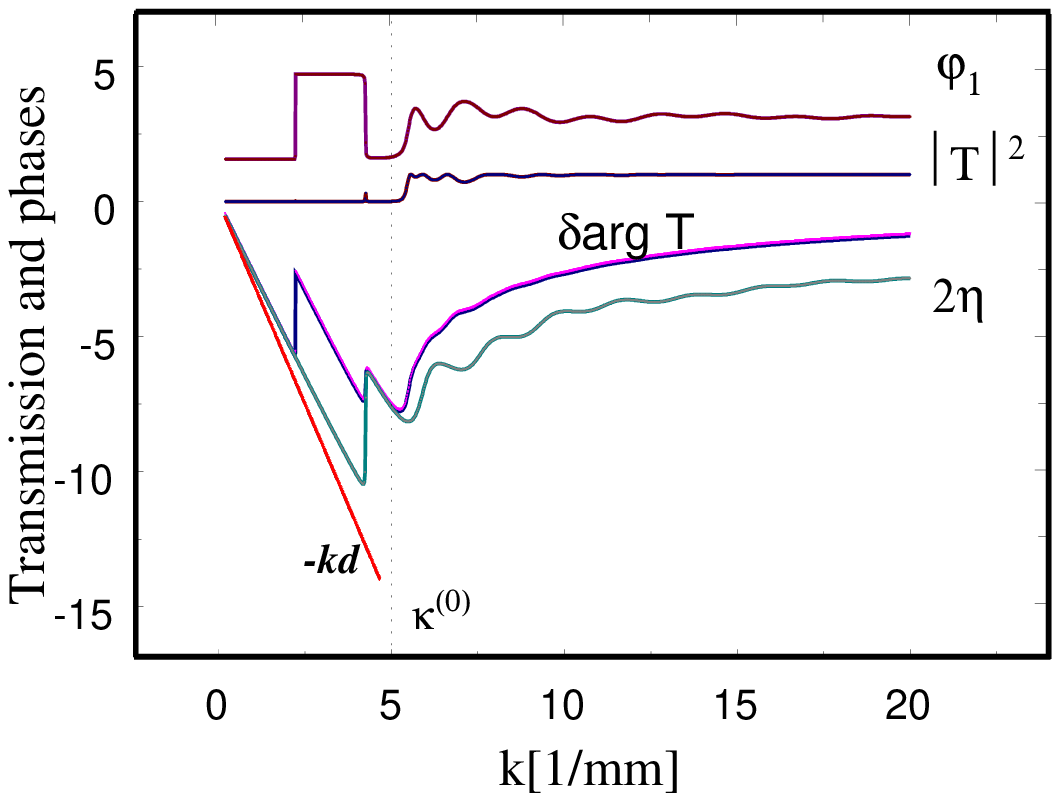,width=90mm,bbllx=120pt,bblly=110pt,bburx=440pt,bbury=350pt}
\end{minipage}

FIGURE 6. Transmission through two equal barriers each one length unit  [mm] wide. The cavity diameter is one 
millimeter wide too. $\kappa^{(0)}$ is barrier height in $k$ units [1/mm] the same as for one barrier tunneling 
(cf. fig.4).

We can write general expression for monodromy single cell traces 
\begin{equation}
\begin{array}{l}
 \cos \phi _i  = \frac{1}{2}Tr \left[M_i \right]  = {\mathop{\rm Re}\nolimits} \frac{{\exp ( - i(\varphi _{1,i}^{}  
+ \varphi _{2,i}^{}  ))}}{{\sin (\varphi _{1,i}^{} )}} = \frac{{\cos (\varphi _{1,i}^{}  + \varphi _{2,i}^{}  
)}}{{ T_{1,i} }} \\ 
  = \cot(\varphi _{1,i}^{} )\cos (\varphi _{2,i}^{}  ) - \sin (\varphi _{2,i}^{}  ) \\ 
 \end{array}
\end{equation}
For each cell we can define two internal phases $\varphi_{1,i}$, $\varphi_{2,i}$ "Bloch phases" ( in analogy to   
$\phi_i $ in [10]) and another one $  k \Delta_i$ as external typical for the interbarrier movement. Such method 
can be  compared  to the scattering as in [10]. Stability in classical mechanic is expressed by inequality $TrM \leq 
2$, here in quantum mechanics $TrM$ describes mutual ratio of reflection and transmission (tunnelling) -
expressed respectively by $\phi_i $ or $\varphi_i$ behavior. High above the barriers $\varphi _{1,i} \rightarrow 
\pi /2 $ so $ \phi_i \rightarrow  \varphi _{2,i} \rightarrow kd_i $. We see that much more appropriate phases to 
be named "Bloch phases" are $\varphi _{1,i},\varphi _{2,i}  $ which describe internal device structure however  
here we deal with the aperiodical or quasiperiodical system (periodic +boundary).

Let assume there exist average transfer operator $\underline {M}$ equivalent to superposition of  equal or 
different 
elementary monodromy cells (assuming symmetric case $a_n+\epsilon_n = -a_1 + \epsilon_1$). We say there 
exist an equivalent "black box" barrier operator $\underline{M}$  which preserves the single cell form (cf. eq. 
(1.15,17,18)).

\begin{equation}
\left[ {O_{\Delta _n }^{(U)} } \right]^{-1} \prod\limits_{i = 1}^n { \left[ {O_{\Delta _i }^{(U)} } \right] \left[ 
{H_i^{(U)} } \right] = \underline M  = \left( {\begin{array}{*{20}c}
   {\frac{{e^{i(\underline {\varphi _1^{} }  + \underline {\varphi _2^{} } )} }}{{\sin (\underline {\varphi _1^{} 
} )}}} & { - \cot(\underline {\varphi _1^{} } )}  \\
   { - \cot(\underline {\varphi _1^{} } )} & {\frac{{e^{ - i(\underline {\varphi _1^{} }  + \underline {\varphi 
_2^{} } )} }}{{\sin (\underline {\varphi _1^{} } )}}}  \\
\end{array}} \right)}  = \left( {\begin{array}{*{20}c}
   {f^ {\dag}  (d,\zeta )} & {g^ {\dag} (d,\zeta )}  \\
   {g(d,\zeta )} & {f(d,\zeta )}  \\
\end{array}} \right)
\end{equation}

The phases $\underline{ \varphi _1} , \underline {\varphi_2} $ are functions of device internal structure. We 
assume such phases exist and can be computed while $M$ is folded from square barriers, $Te^{ikd}=f^{-
1}(d,\zeta) = \sin (\underline{\varphi _1^{}}) e^{i(\underline {\varphi _1^{} }  + \underline {\varphi _2^{} })}$ 
where $\zeta$ represents all internal variables needed to compute $ \underline {\varphi _1^{} } ,   \underline 
{\varphi _2^{}} $ . In most cases the product of  the single cell expressions (1.18) can be computed only 
numerically. The total phase shift change is given 
approximately by ($\underline{ M}$ transforms input $(1,R)$ onto $(T,0)$  output):

\[
\underline \varphi  _{Monodromy}  = (\underline {\varphi _1 }  + \underline {\varphi _2 } )  = \arg \{ f(d,\zeta 
)\}  =" tan^{ - 1} ({\textstyle{1 \over 2}}(  {\textstyle{1 \over {\underline \sigma  }}}-\underline \sigma   )\tanh 
(2\underline {\kappa \varepsilon } ))" 
\] 

Equivalent "black box" width is $d=\underline{2\epsilon}$ .  
Well above the barriers  i.e.
$ \arg \{ f(d,\zeta(\sigma) )\} \mathop  \Rightarrow \limits_{\sigma  \to i=\sqrt{-1}} kd $
there is the transmitted wave  running in the initial direction nearly without distortion. Here the argument of  $ 
\tan^{ - 1} $ has only symbolic meaning however we assume existence of $ \underline {\sigma} , \underline 
{\epsilon} , \underline {\kappa} $ for the equivalent black box barrier.
 
In equations (1.25), for periodic case (i.e. $M_i= M_{\varepsilon ,\Delta }$ ) where Cayley –Hamilton theorem 
can be applied,  
$  \left[ {O_{\Delta _n }^{(U)} } \right]^{ - 1}  $ just cancel exponential term on the diagonal  of $M_i$ matrix.
The total transfer periodic matrix is if $\cos(\phi)=\frac{1}{2}Tr M_{\varepsilon ,\Delta }$:

\[
\begin{array}{l}
 M^{} = \left[ {O_\Delta ^{(U)} } \right]^{ - 1} M_{\varepsilon ,\Delta }^N  = \left[ {O_\Delta ^{(U)} } 
\right]^{ - 1} \left[ {M_{\varepsilon ,\Delta }^{} \frac{{\sin (N\phi )}}{{\sin \phi }} - I\frac{{\sin ((N - 1)\phi 
)}}{{\sin \phi }}} \right] =  \\ 
\left[  H_\varepsilon ^{(U)} \right]  \frac{{\sin (N\phi )}}{{\sin \phi }} - \left[ {O_\Delta ^{(U)} } \right]^{ - 1} 
\frac{{\sin ((N - 1)\phi )}}{{\sin \phi }} \\ 
 \end{array}
\]

it is obvious that similarity of $ M_{\varepsilon ,\Delta }^{} $ trace with $ M^{} $ trace (as in [10]) 
is accidentally. The trace properties are useful when we want to write any power of $ M_{\varepsilon ,\Delta 
}^N$ in terms of $ M_{\varepsilon ,\Delta }$ and 
the unit matrix (see ref [13]). There exists however an additional matrix factor $\left[ {O_\Delta ^{(U)} } \right] 
$ which changes the final trace completely. From physical point of view we are interested only in  
transformations which put $\left[ {O_\Delta ^{(U)} } \right] $  into diagonal form meaning that between 
interaction areas we have two free waves 
running in opposite directions. In general case for the $i$-cell we can define two scattering or Bloch phases 
$\varphi _{2,i}, \varphi _{1,i}$ (Bloch phases suggest periodicity what is not here the case) and phase 
displacement $k\Delta_i$ so each cell has different trace properties. During out of resonance tunneling, particle 
seems to be insensible to  $\Delta_i $ distances.  

 It is impossible to make diagonal both $\left[ {O_\Delta ^{(U)} } \right]_i $ and  
$\left[  H_\varepsilon ^{(U)} \right]_i $. Matching conditions induce the monodromy form of  $\left[  
H_\varepsilon ^{(U)} \right]_i $ describing particle movement under the barrier as in [6]. The total transfer 
matrix is composed of many matrices. We must know its final form explicitly to find mutual ratio of reflection 
and transmission.

In aperiodical symmetric system (like barriers on the Cantor set)  the averaged term $ \underline{ \varphi _1}  + 
\underline{\varphi _2}  $ is different from the expression  valid for the single barrier (1.21,22) . 
The case symmetric is important, it is easy  to calculate the phases $\varphi_1,\bar \varphi_2$ : in the 
monodromy matrix (1.16) $W=0$ and as written already $V$ is real so from (1.24) $V= \cot 
(\underline\varphi_1)=R/T$. From $X,Y$ we extract the second phase. For asymmetry we get  $\cot 
(\underline\varphi_1)=\pm \sqrt {V^2+W^2}= R/T$ and $ \tan (\Delta \underline\varphi ) = \frac{W}{V}$.
$\Delta \underline\varphi (k)$ is phase difference between amplitudes in reflection and transmission.

Resume:

Translation operators suggest  "translation in time" too. $S$-matrix is time independent operator   
(${U(t\rightarrow -\infty,+ \infty)} $) and treats the quantum wire translation device as "black box" which 
structure should be find out in "phase shifts" experiments. Unitarity of  $S$-matrix suggests full symmetry of  
"black box". It is not clear if $S$-matrix phase shifts $\eta_l$ can be used to calculate time delay for particle 
traveling through the investigated object cf.[1].

 Only the transmitted (tunneling) waves "feel" the object size (i.e. its depth). Tunneling introduces asymmetry  
into experiment as well into theory. May be motion emerges in the quantum mechanics as consequence of 
reflection - transmission interference (which takes place only under / over the barrier or the potential well). So 
there exist equivalent $S$-matrix ($S_M$) related with $M$-monodromy  translation operator if $M$ is 
symmetric. Asymmetry causes problems and microreversibility is exact up to tunneling. $M$-matrix can be 
periodic but generally it is not. 
To describe properly tunneling and reflection we need mixture of inward and outward solutions (or at each point 
of the space the true solution is mixture of regular and irregular one coupled by tunneling effect at origin where 
integration of the wave equation starts).

In tunneling, due to equivalence between the complex Schr\"{o}dinger and two dimensional Maxwell 
(Helmholtz) equations, we consider as well photons as massive particles. 

\subsection{Monodromy time}

Two phase shifts suggest that Wigner causality should be applied separately to reflected and transmitted waves.
But if reflection has nothing common with transmission their coupling through the $|R|^2 +|T|^2=1$ relation 
would result  in completely different phases of both functions. The above analysis shows that in principle for 
symmetric systems both waves have common phase $\bar{\varphi}_2=\varphi_1 +\varphi_2-kd$ [8] and such 
sum should be used in causality relation. The phase $\varphi_2$ alone corresponds to $\eta$ ($\eta_l$) if $ k \gg 
\kappa^{(0)}$ where  $\kappa^{(0)}$ is the barrier height i.e. $\sqrt {2mV/\hbar^2}$. 
We can generalize the Wigner time and write
\begin{equation}
\begin{array}{cc}   
t_{ Monodromy } =\frac{\partial (\bar{\varphi}_2+dk)}{\upsilon \partial k} > 0 ; &  \bar{\varphi}_2=\varphi_1 
+2\eta 
\end{array}
\end{equation}
Tangent to  $\varphi_{Monodromy}=\bar{\varphi}_2(k)+dk$ cannot be negative: the scattered wave cannot 
leave the barrier of width $d$ before the incoming wave has entered it - in consequence $ t_{ Monodromy }>0 $. 
There is additional phase $\varphi_1$ which modifies weak causality relation (eq. 1.11). Both phases result from 
$M$-matrix.

In case of asymmetric barrier systems we can introduce times separately for reflection and transmission:
\begin{equation}
\begin{array}{cc}   
t_{ Monodromy,refl } =\frac{\partial (\bar{\varphi}_2+dk)}{\upsilon \partial k}  ; &  t_{ Monodromy,trns } 
=\frac{\partial (\bar{\varphi}_2+dk+\Delta\varphi)}{\upsilon \partial k} 
\end{array}
\end{equation}

\section{ Nimtz experiments in view of the monodromy matrix}

\setcounter{equation}{0} 

The barriers in Nimtz experiment [12] consist of two photonic lattices which are separated by an air gap. Each 
lattice consists between one and four equidistant Perspex layers separated by an air. The refractive index of 
Perspex is n=1.61 in the measured frequency region. In order to build a photonic barrier for the microwave 
signals, the thickness of the Perspex b=6.0(or 5.0) mm and the air layers a=12.0, (8.5) mm present a quarter of 
the microwave carrier's  wavelength in barrier $\lambda_n =c/(nf_c)=22.1, (20.4) mm$ and in air  $\lambda_0 
=c/f_c =35.5, (32.8) mm$ respectively. The air space $d_{cav}=130, (189) mm$ between the two lattices forms 
a cavity and extends the total length of the barrier. The resonance frequencies of the cavity can be calculated on 
the base of monodromy matrix and are in case of two setups  1097.Mhz (or 764.MHz); according to Nimtz 
$f_{res}=c/(2d) $ is (1153) or (794MHz).

The calculated transmission and wave function phases (according to monodromy for symmetric photonic lattices 
the reflection phase equals the transmission phase) are displayed in Fig(7,8). In Fig 8 we marked three areas with 
anomalous dispersion. 

Nimtz assumes that the frequency spectrum of the microwave signal lies completely in the nonresonant 
"forbidden" frequency region between $11f_{res}$ and $12f_{res}$. Using the monodromy calculation method 
and if $\kappa^{(0)}$ is correct, $f_{c}$  should be shifted in comparison to Nimtz data (see figs 7,8) i.e. $k$ 
from the value 0.1769 up to $\sim 0.19$ equivalent , $f_{c'}\simeq 9GHz$   . The superluminal k-regions weakly 
depend on small $\kappa^{(0)}$ changes. However $\kappa^{(0)}$ should be determine from the internal 
Perspex structure. Sometimes for one barrier there is the sharp change in both $\varphi_1 , \bar \varphi_2 $ 
phases behaviour at $ k=\kappa^{(0)}$ if barrier width is comparable or bigger then particle wave length .

for 

\begin{minipage}{100mm}
\epsfig{file=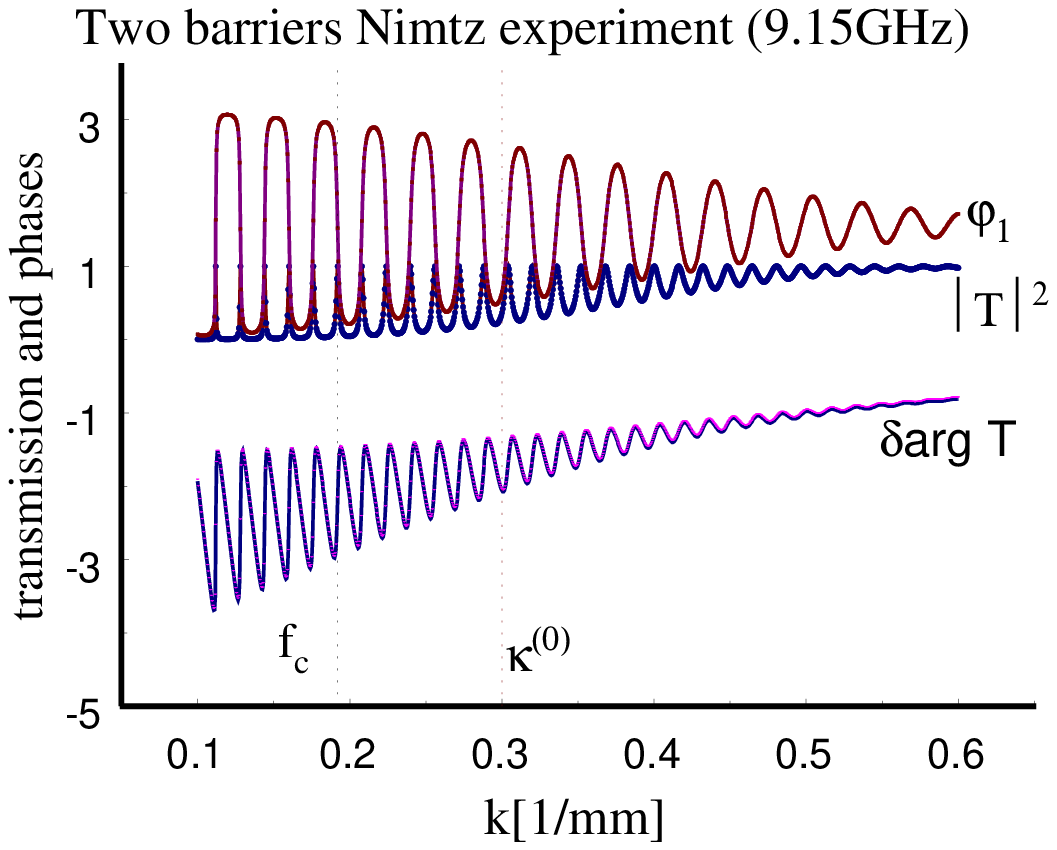,width=90mm, bbllx=120pt,bblly=110pt,bburx=440pt,bbury=370pt }
\end{minipage}

FIGURE 7. Two barriers (9.15GHz) experiment: thickness of barrier is $5.0mm$, $d_{cav}=189mm$. The 
superluminal speed changes gradually from $6.9c$ at $k=0.12$ to $3.4c$ near $f_{c}$, cf. fig8. Total width of 
the system is 199mm.

\begin{minipage}{100mm}
\epsfig{file=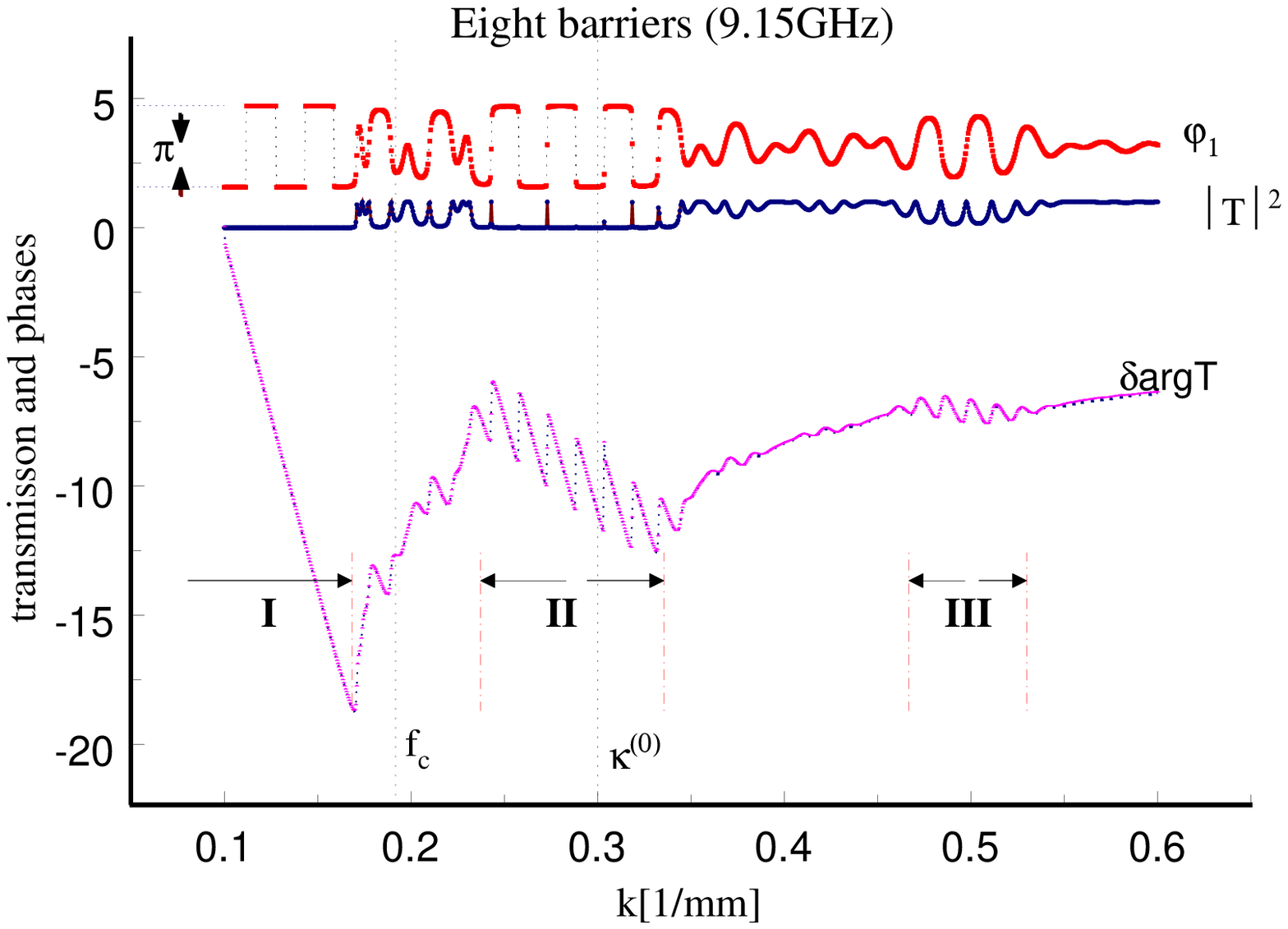,width=90mm,bbllx=50pt,bblly=70pt,bburx=520pt,bbury=420pt}
\end{minipage}                                             

FIGURE 8. Eight barriers (9.15GHz) experiment at $9.15GHz$: thickness of barrier is $5.0mm$, of air layers is 
$8.5mm$, while $d_{cav}=189mm$. The superluminal  speed in region $I$ is $\sim23c$ while in region $II$ 
$\sim 14c$. Total width of the system is 280mm.

It is not easy, from the phase curves to say where $\kappa^{(0)}$ is placed. $\kappa^{(0)}$ should be found by 
any independent method.

At the end we present phase shift analysis in case of the Kiang model with 10 $\delta$-barriers [10,11]. In 
allowed bands the "quantum hurdler" is retarded. But it is not so simple in case of forbidden bands. If the particle 
is reflected, negative slope can be related with penetration depth and for the transmitted wave with advance 
speed.

\begin{minipage}{100mm}
\epsfig{file=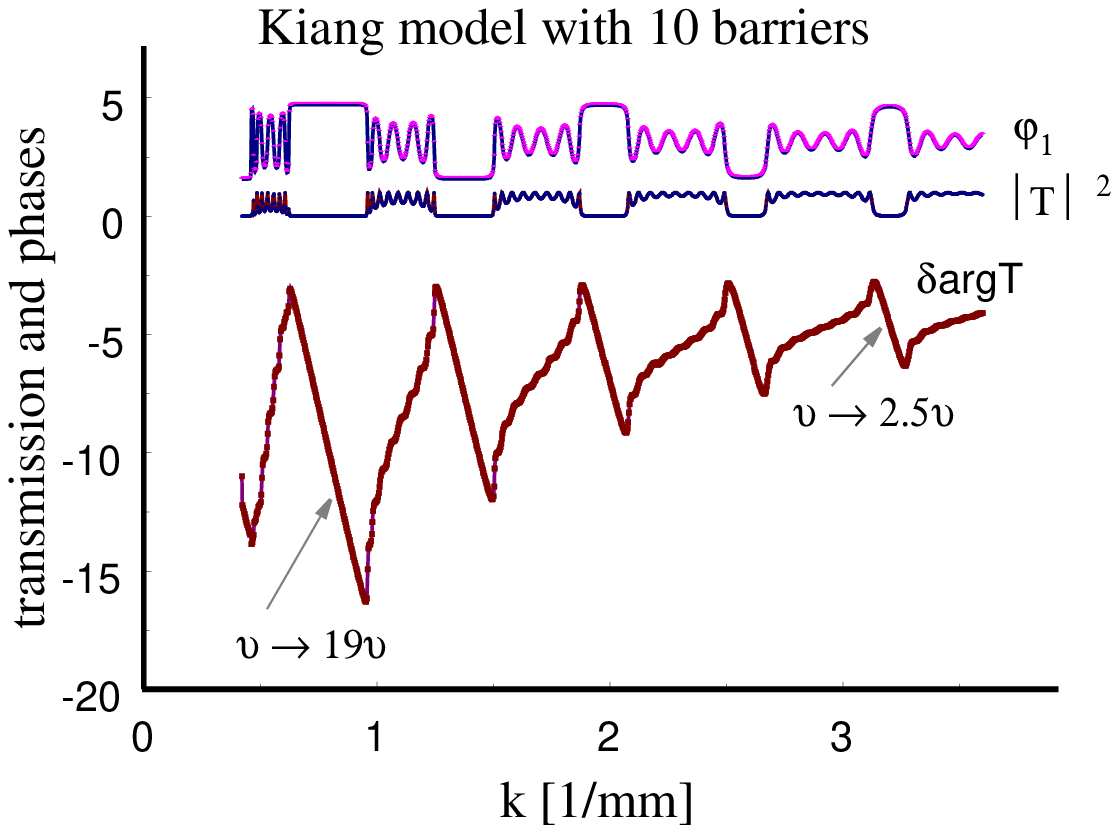,width=90mm,bbllx=110pt,bblly=110pt,bburx=450pt,bbury=360pt}
\end{minipage}                                             

FIGURE 9.  The Kiang model with ten $\delta$ barriers is a good example of  ("superluminal") time advance  
effect in the successive k-forbidden regions. As in [10], we used for that model $\Omega \Delta_{\epsilon}=5$, 
where $\Omega$ is barrier penetrability and $\Delta_{\epsilon}$ is the interbarrier distance. The slope of the 
phase line characterizes the superluminal speed i.e. for the deepest band we have $\sim 19c$ for the next bands 
this speed decreases and is adequately: 10c, 5.5c, 4c and 2.5c in the latest band. In allowed regions the slopes are 
positive and give retardations. The group time delay is much bigger at resonances then between them - the phase 
curve oscillates strongly.

\subsection{ Final remarks}

We assume that $d$ is system width and $\tau^{(0)}=d/(c)$ is the "classical" time needed to travel through 
barriers. For symmetric device both waves – reflected and transmitted – received the same phase shift 
$\bar{\varphi}_2$ (modulo $\pi/2$). The monodromy phase shift analysis of reflection and transmission 
amplitudes rejects reflection from the front and treats both processes as occurring after the time $ 
\tau_{tun}={\tau ^{(0)} + \tau _{}^{(2)} } $ where $\tau^{(2)}$ results from $\bar\varphi_2$ . From the $M$- 
matrix point of view  reflection and transmission  delay  if  defined as  $\tau_{tun} = (d+\partial 
\bar\varphi_2/\partial k)/\upsilon = 2d_{pen}/\upsilon $ is positive and allow us to interprete  $ d_{pen} $ as 
penetration depth in case of reflection. But we are unable to say from which position particle is transmitted.  In 
that case interpretation is not easy. May be the tunneling wave function is strongly repelled from forbidden bands 
resulting in negative phase derivative. It seems that the tunneling particle needs much shorter time to travel 
through barrier than in free space. Numerical calculations (with M-matrix) show that always  $ d_{pen}>0 $ 
according to weak Wigner causality. If $2d_{pen}$ is actual distance seeing by the tunneling particle then  
$2d_{pen}/\tau_{tun}$ is typical speed ($\upsilon$) in the matter, but $d/ \tau_{tun}$ gives  the advance speed. 
We stress once more the transfer matrix enable us to find $k$-dependence of phases.
The phase shift $\varphi_2$ alone is not monotonic and corresponds to phase shifts from S matrix. Analysis of 
amplitudes from equivalent $S$ matrix cause problems. We must define the arrival time for reflected particle and 
"departure" time for transmitted one. The probabilistic interpretation of R and T suggests that reflection occurs at 
barrier front [14] and then anomalous dispersion brakes weak Wigner causality, $\tau^{(0)}$ time must be 
separately defined for the reflected and transmitted waves, in consequence there is no interference between these 
waves. 
 When we calculate $\bar \varphi_2(k)$ in allowed bands in simple $\delta$-barrier systems we recognize that it 
is monotonic function of $k$ with nontypical resonance structure of another origin. The quantum hurdler is 
retarded as it can be seen from $\delta \arg T$ in the Kiang model (fig. 9) but forbidden bands push out the 
particle. This structure emerges from interference effects between both ( reflected and transmitted or incoming 
and reflected) elastic channels in "continuum" including tunneling. Possible superluminal area are seen in Fig.(5-
9). i.e. anomalous dispersion  $\frac{\partial \bar  \varphi_2}{\partial k}<0$ but it is questionable if 
$\varphi_1=const$ or is negligible - $\varphi_1$ is oscillating function of wave number $k$ . We analyze 
anomalous dispersion (without absorption or dissipation) only from two channel interaction point of view. In 
usual propagation of light in refractive media we reject influence of reflected wave.

Brillouin has written in his book [15] "it is impossible to think of refractive medium without dispersion" (and 
energy loss) so the questions - what velocity coincides with elastic tunneling as well what the nature of 
anomalous dispersion is - remain open.

We thanks to Nimtz for kind scientific cooperation. For stimulating discussion we are grateful to A.Horzela, 
V.S.Olkhovsky, E.Recami, S.Maydaniuk. 

{\em References}

[1] Smith, F.T, {\it Phys.Rev.} {\bf 118} 349 (1960) ; Ohmura,T. {\it Progress of Theor.Phys.} {\bf 29} 108 
(1964).

[2] Bohm, D.(1951),Quantum Theory, Prentice-Hall, New York.  

[3] Wigner,E,P.,{\it Phys. Rev.} {\bf 98} 145 (1955).

[4] Olkhovsky,V.S., Recami,E. {\it Phys. Rep.} {\bf 214} 340 (1992).

[5] Joachain,Ch.J., .,{\it Quantum  Collision Theory },North Holland Publishing  Co., (1975) Ch.4.4;
    Taylor,J.R., {\it Scattering theory } J.Wiley \& Sons., N.Y. London, Sydney, Toronto (1972), Ch.11;

[6] Arnold,V.I. .,{\it Geometric Methods in the Theory of Ordinary Differential Equations },Springer, (1987)

[7] Jaworski,W.  Wardlaw,D.M. {\it Phys.Rev.} {\bf A37} 2843 (1988).

[8] Jakiel,J. Olkhovsky,V.S. Recami,E. {\it Phys.Lett.} {\bf A248} 156 (1998)

[9] Fl\"{u}gge,S. ,{\it Rechenmethoden der Quantentheorie}, Springer-Verlag 1990 

[10] Sprung, D.W.L. Hua Wu, Martorell,J. {\it Am.J.Phys.}{\bf 61} 1118 (1993)

[11] Kiang,D. {\it Am.J.Phys.}{\bf 42} 785  (1974)

[12] Nimtz,G. {\it Nonlocal reflection by photonic barriers},  CERN {\it physics/0103073,  /0204043}

[13] Vezzetti,D.J. Cahay,M.M. {\it J.Phys.D} {\bf 19} L53 (1986)

[14] Hauge,E.H.  St{\o}vneng,J.A.{\it Rev. of Mod.Phys.} {\bf 61} 917 (1989) 

[15] Brillouin,L. {\it Wave propagation and group velocity} Academic Press New York (1960)

[16] Peierls,R. {\it Surprises in theoretical physics } Princeton (1979)


\newpage

{\bf Figure descriptions} 

FIGURE 1. Inward,outward, tunneling etc. waves in scattering. During scattering only one wave (here) outward 
or inward is modified. \underline{ There is no cross terms between inward and outward fluxes}. $\Psi_{tun}$ is 
not incorporated in $\Psi^{outw}$ nor in $\Psi^{inw}$. It is not clear if the reflected wave $\Psi_{refl}(:|S|=1)$ 
is equal to $\Psi^{outw}$. In shadow region the complete wave function must vanish [16] ($\psi=0$), there is no 
place for $\Psi_{tun}$ in $\psi$.  

FIGURE 2. {The monodromy problem as defined in [6] for one barrier. Mutual relations between ingoing and 
outgoing  (from left or right side) particle wave functions are displayed. (In the picture bars mean complex 
conjugations.) The transition from initial state  to final one $\textstyle{ \left( {\begin{array}{*{20}c}
   {\Phi _{( + )} }  \\
   {\Phi _{( - )} }  \\
\end{array}} \right)_{in}  \to \left( {\begin{array}{*{20}c}
   {\Phi _{( + )} }  \\
   {\Phi _{( - )} }  \\
\end{array}} \right)_{out}} $ is given by the unimodular M matrix  $\textstyle{ \left( {\begin{array}{*{20}c}
  {1/\bar T} &    {\bar R/\bar T} \\
{R/T}  & {1/T}   \\
\end{array}} \right)} $ easy deducted from transitions as drawn in picture. }

FIGURE 3.     Set of barriers as used in tunneling and transmission or reflection. 

FIGURE 4. Transmission through one barrier three units [mm] wide. $\kappa^{(0)}$ is barrier height in $k$ 
units [1/mm]. $-kd$ represents maximal negative phase slope according to weak Wigner causality. Slope of  
$\bar{\varphi}_2=\varphi_2 + \varphi_1-kd$ is connected with the group velocity in transmission through the 
barrier. $(\partial/\upsilon\partial k) (\bar{\varphi}_2)=\delta \tau$ is time "delay". The wave length $\lambda$ at 
$\kappa^{(0)}$  is smaller than the barrier width.

FIGURE 5. Transmission through one Nimtz barrier 6mm wide. $\kappa^{(0)}$ is barrier height in $k$ units 
(here $mm^{-1}$). As in fig.4 ($-kd$) represents maximal negative phase slope according to weak Wigner 
causality. The wave length $\lambda$ at $\kappa^{(0)}$  is bigger than the barrier width so phase characteristic 
is dominated by $\varphi_1$. The slopes of $\varphi_1(k)$ and $\delta \arg{T}$  phase curves  are positive and 
give retardation. However $\varphi_2-kd=2\eta$ suggests small speed advance.

FIGURE 6. Transmission through two equal barriers each one length unit  [mm] wide. The cavity diameter is one 
millimeter wide too. $\kappa^{(0)}$ is barrier height in $k$ units [1/mm] the same as for one barrier tunneling 
(cf. fig.4).

FIGURE 7. Two barriers (9.15GHz) experiment: thickness of barrier is $5.0mm$, $d_{cav}=189mm$. The 
superluminal speed changes gradually from $6.9c$ at $k=0.12$ to $3.4c$ near $f_{c}$, cf. fig8. Total width of 
the system is 199mm.

FIGURE 8. Eight barriers (9.15GHz) experiment at $9.15GHz$: thickness of barrier is $5.0mm$, of air layers is 
$8.5mm$, while $d_{cav}=189mm$. The superluminal  speed in region $I$ is $\sim23c$ while in region $II$ 
$\sim 14c$. Total width of the system is 280mm.

FIGURE 9.  The Kiang model with ten $\delta$ barriers is a good example of  ("superluminal") time advance  
effect in the successive k-forbidden regions. As in [10], we used for that model $\Omega \Delta_{\epsilon}=5$, 
where $\Omega$ is barrier penetrability and $\Delta_{\epsilon}$ is the interbarrier distance. The slope of the 
phase line characterizes the superluminal speed i.e. for the deepest band we have $\sim 19c$ for the next bands 
this speed decreases and is adequately: 10c, 5.5c, 4c and 2.5c in the latest band. In allowed regions the slopes are 
positive and give retardations. The group time delay is much bigger at resonances then between them - the phase 
curve oscillates strongly.

\end{document}